\documentclass[12pt,a4paper]{article}

\usepackage{amsmath}
\usepackage{amsfonts, amssymb}
\usepackage{amscd}

\newcommand{\EQ}{\begin{equation}}
\newcommand{\EN}{\end{equation}}

\newtheorem{theo}{Theorem}
\newtheorem{defi}{Definition}
\newtheorem{coro}{Corollary}
\newtheorem{prop}{Proposition}
\newtheorem{lemm}{Lemma}

\newtheorem{ex}{Example}

\newcommand{\pr}{\indent{\em Proof: \ }}
\newcommand{\qed}{\hspace*{5 mm}$\triangle$\bigskip}

\newcommand{\Z}{{\mathbb{Z}}}

\newcommand{\dd}{\displaystyle}

\newcommand{\codi}{{\cal C}}

\newcommand{\zero}{{\mathbf{0}}}
\newcommand{\ad}{\mathbb{Z}_2^{\alpha}\times \mathbb{Z}_4^{\beta}}
\newcommand{\add}{\Z_2\Z_4}
\renewcommand{\u}{{\mathbf{1}}}
\newcommand{\dos}{{\mathbf{2}}}

\newenvironment{demo}{\noindent {\pr}\ }{\qed}
\newcommand{\cH}{\cal{H}}
\newcommand{\cG}{{\cal G}}
\newcommand{\bgamma}{\bar{\gamma}}
\newcommand{\bdelta}{\bar{\delta}}
\newcommand{\bkappa}{\bar{\kappa}}

\title{$\add$-linear codes: generator matrices \\ and duality
\thanks{This work was supported in part by the Spanish MEC and the European FEDER  under Grant MTM2006-03250 and also by the UAB under Grant PNL2006-13. The material in this paper was presented in part at {\it V Jornadas de Matem\'{a}tica Discreta y Algor\'itmica}, Soria, Spain, July 2006.}}

\author{J. Borges; C. Fern\'{a}ndez; J. Pujol; J. Rif\`{a}; M. Villanueva \thanks{The authors are members of the Department of Information and
Communications Engineering, Universitat Aut\`{o}noma de Barcelona, 08193-Bellaterra, Spain.} \\ }

\date{}

\begin{document}

\maketitle

\begin{abstract}
A code ${\cal C}$ is $\add$-additive if the set of coordinates can
be partitioned into two subsets $X$ and $Y$ such that
 the punctured code of $\codi$ by deleting the coordinates outside $X$
(respectively, $Y$) is a binary linear code (respectively, a quaternary linear code). In this paper $\add$-additive
codes are studied. Their corresponding binary images, via the Gray
map, are $\add$-linear codes, which seem to be a very distinguished
class of binary group codes.

As for binary and quaternary linear codes, for these codes the fundamental
parameters are found and standard forms for generator and parity check
matrices are given. For this, the appropriate inner product is deduced and
the concept of duality for $\add$-additive codes is defined. Moreover,
the parameters of the dual codes are computed.
Finally, some conditions for self-duality of $\add$-additive
codes are given.\end{abstract}


\section{Introduction}

Let $\Z_2$ and $\Z_4$ be the ring of integers modulo 2 and 4
respectively. Let $\Z_2^n$ denote the set of all binary vectors of
length $n$ and let $\Z_4^n$ be the set of all quaternary vectors of
length $n$. Any non-empty subset $C$ of $\Z_2^n$ is a binary code
and a subgroup of $\Z_2^n$ is called a {\it binary linear code} or a
{\it $\Z_2$-linear code}. Equivalently, any non-empty subset ${\cal
C}$ of $\Z_4^n$ is a quaternary code and a subgroup of $\Z_4^n$ is
called a {\it quaternary linear code}.

Quaternary linear codes can be viewed as binary codes under the Gray map defined as $\phi(0)=(0,0),\ \phi(1)=(0,1), \ \phi(2)=(1,1),\ \phi(3)=(1,0)$. If $\codi$ is a quaternary linear code, then the binary code $C=\phi(\codi)$ (coordinatewise extended) is said to be a {\em $\Z_4$-linear} code.
The notions of dual code of a quaternary linear code $\codi$, denoted by $\codi^{\perp}$, self-orthogonal code (when $\codi \subseteq \codi^{\perp}$) and self-dual code (when $\codi=\codi^{\perp}$)
are defined in the standard way (see \cite{MacW}) in terms of the usual inner product for quaternary vectors (see \cite{Sole}). Since in general the binary code $C=\phi(\codi)$ is not linear, it need not have a dual. However, the binary code $C_{\perp}=\phi(\codi^\perp)$ is called the {\it $\Z_4$-dual} of $C=\phi(\codi)$.

Since 1994, quaternary linear codes have became significant due to
its relationship to some classical well-known binary codes as the
Nordstrom-Robinson, Kerdock, Preparata, Goethals or Reed-Muller
codes (see \cite{Sole}). It was proved that the Kerdock code and the
Preparata-like code are $\Z_4$-linear codes and, moreover, the
$\Z_4$-dual code of the Kerdock code is the Preparata-like code.
Lately, also some families of quaternary linear codes, called $QRM$
and $ZRM$, related to the Reed-Muller codes have been studied in
\cite{QRM} and \cite{ZRM}, respectively.

\medskip
Additive codes were first defined by Delsarte in $1973$ in terms of
association schemes (see \cite{del}, \cite{lev}). In general, an
additive code, in a translation association scheme, is defined as a
subgroup of the underlying abelian group. In the special case of a
binary Hamming scheme, that is when the underlying abelian group is
of order $2^n$, the only structures for the abelian group are those
of the form $\Z_2^\alpha\times \Z_4^\beta$, with $\alpha +
2\beta=n$. Therefore, the subgroups $\codi$ of
$\Z_2^\alpha\times\Z_4^\beta$ are the only additive codes in a
binary Hamming scheme. In order to distinguish them from additive
codes over finite fields (see \cite{BG}, \cite{Bier}, \cite{BBrou},
\cite{LP}), from now on we will call them {\em $\add$-additive
codes}.

The binary image of a $\add$-additive code under the extended Gray
map defined in Section \ref{sec:Definitions} is called {\it
$\add$-linear code}. In \cite{Adds} and \cite{PropIT}, binary perfect 1-error
correcting codes (or 1-perfect codes) which are $\add$-linear codes
are described and all such 1-perfect codes are characterized. More
examples, such as extended 1-perfect  and Hadamard codes which are
$\add$-linear codes are studied in subsequent papers (see
\cite{ExtPerf}, \cite{Kro01}, \cite{Hadam}, \cite{PheRif}).  Some notorious codes, e.g.
Kerdock-like and Preparata-like codes, can have a $\Z_4$-linear
structure (see \cite{Sole}), but they cannot have a $\Z_2\Z_4$-linear
structure with non-empty binary part (see \cite{PrepKerd}).

\medskip
As we have seen, the $\add$-additive codes belong to the more
general family of additive codes. However, notice that one could
think of other families of codes with an algebraic structure that
also include the $\add$-additive codes; such as mixed group codes
and translation invariant propelinear codes.

{\em Mixed group codes} are defined as subgroups of a group of type
$G=G_1\times\cdots\times G_r$, where $G_i$ is a finite abelian group
for all $i=1,\ldots,r$ (see \cite{Hede}, \cite{Lind}).
Since any finite abelian group has a
factorization in cyclic groups, we can also think about a mixed
group code as a subgroup of $\Z_{i_1}\times\cdots\times\Z_{i_s}$,
where the indices $i_1,\ldots,i_s$ are not necessarily different. If
we are interested in a binary version of these codes, we need a
one-to-one mapping $\phi$ from $\Z_{i_j}$ to $\Z_2^m$ (where
$2^m\geq i_j$) for all $j=1,\ldots,s$. In \cite{Codis2k}, it was
shown that the indices $i_1,\ldots,i_s$ must be all even, if we want
to use a Gray map $\phi$ which has the classical property that
$d(\phi(i),\phi(i+1))=1$, where $d(\cdot,\cdot)$ is the Hamming
distance between binary vectors. This Gray map is unique, up to
coordinate permutation, if the binary image is also Hamming
compatible. Moreover, it was also proved that if the binary image of
a subgroup of $\Z_{i_1}\times\cdots\times\Z_{i_s}$, using such Gray
map, is a 1-perfect code, then $i_j\in\{2,4\}$, for all
$j=1,\ldots,s$ (i.e. it is also a $\Z_2\Z_4$-linear code).

{\em Translation invariant propelinear codes} were first defined in 1997
(see \cite{PropIT}, \cite{Propes}). In \cite{PropIT}, it was also proved that all such
binary codes are group-isomorphic to subgroups of
$\Z_2^\alpha\times\Z_4^\beta\times Q_8^k$, where $Q_8$ is the
non-abelian quaternion group on eight elements. Hence, abelian
translation invariant propelinear codes are exactly all the
$\Z_2\Z_4$-linear codes.

\medskip
Most of the concepts on $\add$-additive codes described in this
paper have been implemented by the authors as a new package in {\sc
Magma} (see \cite{Magma}). {\sc Magma} is a software package
designed to solve computationally hard problems in algebra, number
theory, geometry and combinatorics. Currently it supports the basic
facilities for linear codes over integer residue rings and Galois
rings; moreover, it also supports functions for additive codes over
a finite field, which are a generalization of the linear codes over
a finite field (see \cite[Chapter 119, 120]{M1}). However, it does
not include functions to work with $\add$-additive codes. For this
reason, a beta version of this new package for $\add$-additive codes
and the manual with the description of all functions can be
downloaded from the web page {\tt http://www.ccg.uab.cat}. For any
comment or further information about this package, you can send an
e-mail to {\em support-ccg@deic.uab.cat}.

\medskip
The aim of this paper is a general study of $\add$-additive codes
and the corresponding $\add$-linear codes. It is organized as
follows. In Section \ref{sec:Definitions}, we give the definition of
$\add$-additive and $\add$-linear codes, we find which are the
fundamental parameters, and we also discuss about the automorphism
groups of these codes. In Section \ref{sec:matrices}, we deduce a
standard form for generator matrices of $\add$-additive codes.
Section \ref{sec:duality} is devoted to study duality for
$\add$-additive codes defining the appropriate inner product. In
Section \ref{sec:parityCheck}, we show how the generator and parity
check matrices are related and we also compute the parameters of the
dual code. Finally, in Section \ref{sec:selfdual}, we give some
conditions for self-duality.

\section{Definitions}
\label{sec:Definitions}

From~now on, we will focus on $\add$-additive codes $\codi$, which
are subgroups of $\Z_2^{\alpha}\times\Z_4^{\beta}$. We will take an
extension of the usual Gray map: $\Phi:
\Z_2^{\alpha}\times\Z_4^{\beta} \longrightarrow \Z_2^{n}$, where
$n=\alpha+2\beta$, given by
$$\begin{array}{lc}
\Phi(x,y) = (x,\phi(y_1),\ldots,\phi(y_\beta))\\
 \hspace{1truecm}\forall x\in\Z_2^\alpha,\;\forall y=(y_1,\ldots,y_\beta)\in \Z_4^\beta;
\end{array}$$
where $\phi:\Z_4 \longrightarrow \Z_2^{2}$ is the usual Gray map,
that is, $$\phi(0)=(0,0),\ \phi(1)=(0,1), \ \phi(2)=(1,1), \
\phi(3)=(1,0).$$ This Gray map is an isometry which transforms Lee
distances defined in the $\add$-additive codes $\codi$ over
$\Z_2^{\alpha}\times\Z_4^{\beta}$ to Hamming distances defined in
the binary codes $C=\Phi(\codi)$. Note that the length of the binary code
$C$ is $n=\alpha+2\beta$.

\medskip
Since $\codi$ is a subgroup of  $\Z_2^{\alpha}\times
\Z_4^{\beta}$, it is also isomorphic to an abelian structure like
$\Z_2^{\gamma}\times \Z_4^{\delta}$. Therefore, $\codi$ is of type $2^\gamma 4^\delta$ as a group, it has
$|\codi|=2^{\gamma+2\delta}$ codewords and the number of order two codewords
in $\codi$ is $2^{\gamma+\delta}$.

Let $X$ (respectively $Y$) be the set of $\Z_2$ (respectively
$\Z_4$) coordinate positions, so $|X|=\alpha$ and $|Y|=\beta$. Unless otherwise stated, the set $X$ corresponds to the first $\alpha$ coordinates and $Y$ corresponds to the last $\beta$ coordinates. Call
$\codi_X$ (respectively $\codi_Y$) the punctured code of $\codi$ by deleting the coordinates outside $X$
(respectively $Y$). Let $\codi_b$ be the subcode of
$\codi$ which contains all order two codewords and let $\kappa$ be
the dimension of $(\codi_b)_X$, which is a binary linear code. For the case
$\alpha=0$, we will write $\kappa=0$.

Considering all these parameters, we will say that $\codi$
(or equivalently $C=\Phi(\codi)$) is of type
$(\alpha,\beta;\gamma,\delta;\kappa)$. Notice that $\codi_Y$ is a
quaternary linear code of type $(0,\beta;\gamma_Y,\delta;0)$, where
$0\leq \gamma_Y \leq \gamma$, and $\codi_X$ is a binary linear code of
type $(\alpha,0;\gamma_X,0;\gamma_X)$, where $\kappa \leq \gamma_X \leq \gamma$.

\begin{defi} Let $\codi$ be a $\add$-additive code, that is a
subgroup of $\Z_2^{\alpha}\times \Z_4^{\beta}$. We say that the
binary image $C=\Phi(\codi)$ is a {\em $\add$-linear code} of
length $n=\alpha+2\beta$ and type
$(\alpha,\beta;\gamma,\delta;\kappa)$, where $\gamma$, $\delta$ and $\kappa$  are defined as above.
\end{defi}

Note that $\Z_2\Z_4$-linear codes are a generalization of
binary linear codes and $\Z_4$-linear codes. When $\beta=0$, the
binary code $C=\codi$ corresponds to a binary linear code. On the
other hand, when $\alpha=0$, the $\add$-additive code $\codi$ is a
quaternary linear code and its corresponding binary code
$C=\Phi(\codi)$ is a $\Z_4$-linear code.

\medskip Two $\add$-additive codes $\codi_1$ and $\codi_2$ both of type
$(\alpha,\beta;\gamma,\delta;\kappa)$ are said to be {\it
monomially equivalent}, if one can be obtained from the other by permutating
the coordinates and (if necessary) changing the signs of certain
$\Z_4$ coordinates. Two $\add$-additive codes are said to be {\it
permutation equivalent} if they differ only by a permutation of
coordinates. The {\it monomial automorphism group} of a
$\add$-additive code $\codi$, denoted by  $MAut(\codi)$,
is the group generated by all permutations and sign-changes
of the $\Z_4$ coordinates that preserve the set of codewords of $\codi$, while
the {\it permutation automorphism group} of $\codi$, denoted by $PAut(\codi)$,
is the group generated by all permutations that preserve the set of codewords of $\codi$ (see \cite{Pless}).

If two $\add$-additive codes $\codi_1$ and $\codi_2$ are monomially
equivalent, then, after the Gray map, the corresponding
$\add$-linear codes $C_1=\Phi(\codi_1)$ and $C_2=\Phi(\codi_2)$ are
isomorphic as binary codes. Note that the inverse statement is not
always true.

\section{Generator matrices of $\add$-additive codes}
\label{sec:matrices}

Let $\codi$ be a $\add$-additive code. Although $\codi$ is not a
free module, every codeword is uniquely expressible in the form  $$\dd
\sum_{i=1}^{\gamma}\lambda_iu_i+\sum_{j=\gamma+1}^{\gamma+\delta}\mu_jv_j,$$
where $\lambda_i \in \Z_2$ for $1\leq i\leq \gamma$,
$\mu_j\in\Z_4$ for $\gamma+1\leq j\leq \gamma+\delta$ and $u_i,
v_j$ are vectors in $\Z_2^\alpha\times
\Z_4^\beta$ of order two and order four, respectively. The vectors $u_i,v_j$ give us a
generator matrix $\cG$ of size $(\gamma+\delta)\times (\alpha+\beta)$ for
the code $\codi$. Moreover, we can write $\cG$ as \EQ \label{eq:matrixG}
    \cG= \left (\begin{array}{c|c} B_1&2B_3\\ \hline B_2&Q\end{array}\right ),
\EN %
where $B_1,B_2,B_3$ are matrices over $\Z_2$ of size $\gamma\times
\alpha$,  $\delta\times \alpha$ and $\gamma\times \beta$,
respectively; and $Q$ is a matrix over $\Z_4$ of size
$\delta\times \beta$ with quaternary row vectors of order four.

In \cite{Sole}, it was shown that any quaternary linear
code of type $(0,\beta;\gamma,\delta;0)$ is permutation equivalent
to a quaternary linear code with a generator matrix of the form
\EQ \label{eq:QaryStandardForm} \cG_S=\left ( \begin{array}{|ccc}
2T & 2I_{\gamma} & \zero\\ \hline
S & R & I_\delta \end{array} \right ), \EN where $R,T$ are
matrices over $\Z_2$ of size $\delta\times\gamma$ and
$\gamma\times(\beta-\gamma-\delta)$, respectively; and $S$ is a
matrix over $\Z_4$ of size $\delta\times(\beta-\gamma-\delta)$.
In this section, we will generalize this result for $\add$-additive codes, so
we will give a canonical generator matrix for these codes (see \cite{BF06}).

First, notice
that changing ones by twos in the coordinates over $\Z_2$, we can
see the $\add$-additive codes as quaternary linear codes. Let
$\chi$ be the map from $\Z_2$ to $\Z_4$, which is the usual
inclusion from the additive structure in $\Z_2$ to $\Z_4$:
$\chi(0)=0, \ \chi(1)=2.$ This map can be extended to the map
$(\chi, Id): \Z_2^\alpha\times\Z_4^\beta \longrightarrow
\Z_4^\alpha\times\Z_4^\beta$, which will also be denoted by
$\chi$.

\medskip
\begin{theo}\label{prop:StandardForm}
  Let $\codi$ be a $\add$-additive code of type
  $(\alpha,\beta;\gamma,\delta;\kappa)$. Then, $\codi$ is
  permutation equivalent to a $\add$-additive code with canonical generator matrix
  of the form
 \EQ \label{eq:StandardForm}
\cG_S= \left ( \begin{array}{cc|ccc}
I_{\kappa} & T_b & 2T_2 & \zero & \zero\\
\zero & \zero & 2T_1 & 2I_{\gamma-\kappa} & \zero\\
\hline \zero & S_b & S_q & R & I_{\delta} \end{array} \right ),
\EN \noindent where $T_b, T_1, T_2, R, S_b$ are matrices over $\Z_2$
and $S_q$ is a matrix over $\Z_4$.
\end{theo}

\begin{demo} Since $\kappa$ is the dimension of the matrix $B_1$ over $\Z_2$
given in (\ref{eq:matrixG}), the code $\codi$ has a generator matrix of the form
$$ \left (
\begin{array}{cc|c}
I_{\kappa} & \bar{B_1} & 2\bar{B_3}\\
\zero & \zero & 2\bar{B_4}\\ \hline
\vspace{-.35truecm} & &  \\
\zero & \bar{B_2} & \bar{Q}
\end{array} \right),
$$ %
where $\bar{B_1},\bar{B_2},\bar{B_3}$ and $\bar{B_4}$ are matrices
over $\Z_2$ of size $\kappa\times(\alpha-\kappa)$, $\delta\times(\alpha-\kappa)$,
$\kappa\times\beta$ and $(\gamma-\kappa)\times\beta$, respectively;
and $\bar{Q}$ is a matrix over $\Z_4$
of size $\delta\times\beta$.

The quaternary linear code $\codi^{-}$ of type
$(0,\alpha-\kappa+\beta;\gamma-\kappa,\delta;0)$ generated by
the matrix $$ \left (\begin{array}{|cc}
\zero & 2\bar{B_4}\\ \hline
\vspace{-.35truecm} &  \\
2\bar{B_2} & \bar{Q}
\end{array} \right) $$
is permutation equivalent to a quaternary linear code with
generator matrix of the form $$\cG^{-}=\left(
\begin{array}{|cccc}
\zero & 2T_1 & 2I_{\gamma-\kappa} & \zero\\ \hline
2S_b &  S_q & R & I_\delta \end{array} \right ),$$ where the
permutation of coordinates fixes the first $\alpha-\kappa$
coordinates (see \cite{Sole} or (\ref{eq:QaryStandardForm})).
So, the quaternary linear code $\chi(\codi)$ generated by the matrix
$$ \left (
\begin{array}{|ccc}
2I_{\kappa} & 2\bar{B_1} & 2\bar{B_3}\\
\zero & \zero & 2\bar{B_4}\\ \hline
\vspace{-.35truecm} & &  \\
\zero & 2\bar{B_2} & \bar{Q} \end{array} \right)
$$ %
is permutation equivalent to a quaternary linear code with
generator matrix of the form $$\cG_\chi= \left (
\begin{array}{|ccccc}
2I_{\kappa} & 2T_b & 2T_2 & \zero & \zero\\
\zero & \zero & 2T_1 & 2I_{\gamma-\kappa} & \zero\\ \hline
\zero & 2S_b & S_q & R & I_{\delta} \end{array} \right ).$$

Finally, $\codi$ is permutation equivalent to a
$\add$-additive code with generator matrix $\chi^{-1}(\cG_\chi)=\cG_S$.
\end{demo}

\begin{ex} \label{example:C1} Let ${\cal C}_1$ denote the
$\add$-additive code of type $(1,3;1,2;1)$ with generator matrix
  $$\cG= \left (\begin{array}{c|c c c} 1 & 2 & 2 & 2\\ \hline
      0 & 1 & 1 & 0\\
      1 & 1 & 2 & 3
\end{array}\right ).$$ The code $\codi_1$ can also be generated by the matrix
$$\left (\begin{array}{c|c c c} 1 & 2 & 2 & 2\\ \hline
      0 & 1 & 1 & 0\\
      0 & 1 & 0 & 3
\end{array}\right ).$$
The quaternary linear code $\codi^{-}$ generated by $\left (\begin{array}{|c c c} \hline
                                              1 & 1 & 0\\
                                              1 & 0 & 3
\end{array}\right )$ is permutation equivalent to a quaternary linear code with
generator matrix $\cG^{-} =\left (\begin{array}{|c c c} \hline
                                               1 & 1 & 0\\
                                               3 & 0 & 1
\end{array}\right ).$
So, the quaternary linear code $\chi(\codi)$ generated by
$$\left (\begin{array}{|c c c c} 2 & 2 & 2 & 2\\ \hline
                                             0 & 1 & 1 & 0\\
                                             0 & 1 & 0 & 3
\end{array}\right )$$ is permutation equivalent to a quaternary
linear code with generator matrix
$$\cG_{\chi}=\left (\begin{array}{|c c c c} 2 & 2 & 0 & 0\\ \hline
                                 0 & 1 & 1 & 0\\
                                 0 & 3 & 0 & 1
\end{array}\right ).$$ Therefore, the code ${\cal C}_1$ is
permutation equivalent to a $\add$-additive code with canonical generator
matrix $$\cG_S=\chi^{-1}(\cG_{\chi})=\left (\begin{array}{c|ccc} 1 & 2 & 0 & 0\\
\hline
                                     0 & 1 & 1 & 0\\
                                     0 & 3 & 0 & 1
\end{array}\right ).$$
\end{ex}

\begin{ex} \label{example:C5} Let $\mathcal{C}_2$ be a
$\add$-additive code of type $(3,4;3,1;3)$ with generator matrix
$$\left (\begin{array}{ccc|cccc}1 & 0 & 0  &  2 & 2 & 0 & 0 \\
                                1 & 1 & 1  &  2 & 2 & 2 & 2 \\
                                1 & 1 & 0  &  2 & 2 & 0 & 0 \\\hline
                                1 & 1 & 1  &  1 & 1 & 1 & 1 \\
\end{array}\right ).$$ 
By Theorem \ref{prop:StandardForm}, $\codi_2$ is
  permutation equivalent to a $\add$-additive code with canonical generator matrix
$$\left (\begin{array}{ccc|cccc}1 & 0 & 0  &  2 & 2 & 0 & 0 \\
                                0 & 1 & 0  &  0 & 0 & 0 & 0 \\
                                0 & 0 & 1  &  2 & 2 & 0 & 0 \\\hline
                                0 & 0 & 0  &  1 & 1 & 1 & 1 \\
\end{array}\right ).$$
\end{ex}

\section{Duality of $\add$-additive codes}
\label{sec:duality}

For linear codes over finite fields or finite rings there exists
the well-known concept of duality. In this section, we will study this
concept for $\add$-additive codes. First, we will show that the inner product of elements
of a finite abelian group can be uniquely defined. Then, considering the finite group
$\Z_2^\alpha \times \Z_4^\beta$ we will define the notions of duality, as the additive dual code and
the $\add$-dual code, for $\add$-additive codes and its corresponding $\add$-linear codes, respectively.

\bigskip
Given a finite abelian group $(G,+)$ of exponent
$m\geq1$ (so, for each $b\in G$ we have $m b=0$), we call {\em dual
group} of $G$, denoted by $\widehat{G}$, the group of ho\-mo\-mor\-phisms
from $G$ into $\Z_m$, $\widehat{G}=Hom(G,\Z_m)$.

\begin{ex}\label{zm} Let $(G,+)$ be the cyclic group of order four
with generator $a$, so $G=\{a,2a,3a,4a=0\}$.
If we know $\varphi(a)$, where $\varphi\in\widehat{G}$, we will know the image
of any element in $G$, because for any $b\in G$ we have $b=i a$
and $\varphi(b)=\varphi(i a)=i \varphi(a)\in \Z_4$. So, there are four
different homomorphisms that we can define over $G$: $$
\begin{array}{r|rrrr}
&\varphi(0)&\varphi(a)&\varphi(2a)&\varphi(3a)\\
\hline
\varphi_0&0&0&0&0\\
\varphi_1&0&1&2&3\\
\varphi_2&0&2&0&2\\
\varphi_3&0&3&2&1 \end{array} $$

Note that, in general, for any cyclic group $(G,+)$ of order $m$
with generator $a$, we can construct all the possible
homomorphisms as $\varphi_k(i a)=k i\in \Z_m$. Note also that
$\varphi_k(b) + \varphi_s (b)=\varphi_{k + s}(b)$, for any $b\in G$.
\end{ex}

It is well-known that ($\widehat{G},\cdot)$ is an abelian group by
using the operation $(\varphi \cdot \lambda )(g)=\varphi(g)+\lambda(g)$, where
$\varphi,\lambda \in \widehat{G}$ and $g\in G$ (see \cite{lang}). The
group $\widehat{G}$ has the same cardinality as $G$ and both
groups are isomorphic (see \cite{lang}), but there is no a
canonical (or natural) isomorphism from $G$ to $\widehat{G}$.

\bigskip Assume $G$ is a cyclic group of order $m$ and fix a
generator $a\in G$. Any homomorphism $\varphi\in
\widehat{G}=Hom(G,\Z_m)$ is defined knowing $\varphi(a)$. If
$\varphi(a)=k\in \Z_m$, this homomorphism will be denoted by $\varphi_k$
and, for any element $b=ia \in G$, $\varphi_k(b)=\varphi_k(i a)=i \varphi_k(a)=i k\in \Z_m$.
So, we can define an isomorphism $G
\longrightarrow \widehat{G}$, such that for all $c\in G$ we
have $c=ja \mapsto \varphi_j$. Note that this isomorphism depends on
the fixed generator $a\in G$.

\bigskip Let $G_1, G_2$ be two abelian groups of exponent $m\geq
1$. A \textit{bilinear map} of $G_1\times G_2$ into $\Z_m$ is a
map $$ \begin{array}{ccc}
G_1 \times G_2 &\longrightarrow & \Z_m\\
(x_1,x_2)&\longmapsto & \langle x_1,x_2\rangle\end{array} $$ such
that for $x_1\in G_1$ the function $x_2 \mapsto \langle
x_1,x_2\rangle$ and for $x_2\in G_2$ the function $x_1 \mapsto
\langle x_1,x_2\rangle$ are homomorphisms.

Let $G$ be an abelian group of exponent $m\geq 1$. A special case of
bilinear map is \EQ \label{bm} G\times \widehat{G} \longrightarrow
\Z_m,\EN where $(b,\varphi) \mapsto \varphi(b) \in \Z_m$, for all
$b\in G$ and $\varphi \in \widehat{G}$. Another special case of
bilinear map is the so called \textit{inner product} in $G$ given by
\EQ \label{bm2} G\times G \longrightarrow \Z_m,\EN where
$(b,c)\mapsto \varphi(b) \in \Z_m$, $\widehat{\varphi}:
G\longrightarrow \widehat{G}$ is a fixed isomorphism and
$\varphi=\widehat{\varphi}(c)$, for all $b,c\in G$.

Note that although the bilinear map given by (\ref{bm})
is canonically defined, the inner product defined by
(\ref{bm2}) depends on the particular isomorphism $\widehat{\varphi}$
from $G$ to $\widehat{G}$ that we use.

\bigskip Assume again $G$ is a cyclic group of order $m$ and fix a
generator $a\in G$. The inner product in $G$ is defined
uniquely by \EQ\label{inn}(b,c) \mapsto \langle b,c\rangle =\varphi_j(b)=\varphi_j(i a)=j i \in \Z_m,\EN
where $b=i a \in G$, $c=j a \in G$ and $\varphi_j \in \widehat{G}=Hom(G,\Z_m)$.

Let $G'$ be a subgroup of $G$ generated by an element $a'\in G'$ of order $t$,
where $t \mid m$.
The dual group of $G'$ could be considered as $Hom(G',\Z_t)$ or
$Hom(G',\Z_m)$ depending on whether the exponent of
$G'$ is $t$ or $m$, respectively.
In both cases a generator $a'$ in $G'$ is send to an
element of order $t$ in $\Z_t$ or $\Z_m$, respectively. This situation can be represented by
$$\begin{array}{ccccc}G' &\longrightarrow &
\Z_t
&\longrightarrow &\Z_m\\
a' & \longmapsto &1 & \longmapsto&\bar{s} \end{array} $$ where
$\bar{s}\in \Z_m$ is an element of order $t$.
Then, after fixing a generator $a' \in G'$ and an element
$\bar{s}\in \Z_m$ of order $t$, the inner product of elements of
$G'$ seen as elements in $G$ is defined uniquely by
$$(b,c) \mapsto \langle b,c\rangle_m = \bar{s} \langle
b,c\rangle_t =\bar{s}
\varphi_j(b)=\bar{s} \varphi_j(i a')=\bar{s} j i \in \Z_m,$$ where $b=i
a'$, $c=j a'$ and $\varphi_j\in Hom(G',\Z_t)$.

\bigskip It is well-known (see~\cite{lang}) that if $G$ is a
finite abelian group, expressed as a product $G=G_1\times G_2$,
then $\widehat{G}$ is isomorphic to $\widehat{G_1}\times
\widehat{G_2}$ under the mapping $\widehat{G_1}\times
\widehat{G_2} \longrightarrow \widehat{G}$, where the element
$(\lambda_1,\lambda_2)\in \widehat{G_1}\times \widehat{G_2}$ is
transformed into an element in $\widehat{G}$ such that for all $(x_1,x_2)\in G$
$$(\lambda_1,\lambda_2)(x_1,x_2)=\lambda_1(x_1)+\lambda_2(x_2).$$

Moreover, any finite abelian group $G$ is isomorphic to a product of cyclic
groups, specifically, we can write $$ G\cong  \Z_{m_1} \times
\Z_{m_2}\times \cdots \times \Z_{m_k},$$ where $ 1 < m_1 \,|\,m_2
\,|\,\cdots \,|\,m_k$.
Therefore, we have that the inner product, defined by
(\ref{inn}) in a cyclic group,
can be extended to any finite abelian group in the
following way:

\begin{prop}\label{defi} Let $(G,+)$ be a finite abelian group of
exponent $m$ and consider the decomposition $$ G\cong  \Z_{m_1}
\times \Z_{m_2}\times \cdots \times \Z_{m_k}, $$ where $1 < m_1
\,|\,m_2 \,|\,\cdots \,|\,m_k=m$ and $m = s_i m_i$,
for all $i=1,\ldots,k$.
After fixing a generator $a_i\in \Z_{m_i}$
in each component and elements $\bar{s_i}\in \Z_{m}$ of order
$m_i$, any element $u=u_1a_1 +u_2a_2+\cdots +u_ka_k$
in $G$ is expressed as $u=(u_1,u_2,\cdots,u_k)\in G$ in this fixed
generators system.

The inner product of elements $u=(u_1,u_2,\cdots,u_k),
v=(v_1,v_2,\cdots,v_k)\in G$ is  defined uniquely by
\EQ\label{innsubf}\langle u,v\rangle_m = \sum_{i} \bar{s_i}
\langle u_i,v_i\rangle_{m_i}  =\sum_{i}\bar{s_i}
\varphi_{v_i}(u_i)=\sum_{i}\bar{s_i} v_i u_i\in \Z_m.\EN \end{prop}

\bigskip
Now, consider the finite abelian group $\Z_2^\alpha \times
\Z_4^\beta$ whose elements are vectors of $\alpha+\beta$ coordinates
(the first $\alpha$ over $\Z_2$ and the last $\beta$ over $\Z_4$).
By Proposition~\ref{defi}, fixing generators $a_i=1\in \Z_2$, for
$1\leq i\leq \alpha$, and $a_i\in\{1,3\}\in \Z_4$,  for
$\alpha+1\leq i\leq \alpha+\beta$, and also fixing the values
$\bar{s_i}=2$, for $1\leq i\leq \alpha$, which is the only possible
value of order two in $\Z_4$, and $\bar{s_i}=1\in\{1,3\}\subset
\Z_4$, for $\alpha+1\leq i\leq \alpha+\beta$, we can write the inner
product given by (\ref{innsubf}) in the following way that we will
call \textit{standard inner product}:

$$ \langle u,v \rangle=2(\sum_{i=1}^{\alpha} u_iv_i)+\sum_{j=\alpha+1}^{\alpha+\beta}
u_jv_j\in \Z_4,$$ where $u,v\in \Z_2^{\alpha}\times \Z_4^{\beta}$.

Note that although  $\bar{s_i}$ is uniquely defined for $1\leq i\leq
\alpha$, the value of $\bar{s_i}$, for $\alpha+1\leq i\leq
\alpha+\beta$, can be chosen from $\{1,3\}$ and so, we can produce
several different presentations for the inner product. Also note
that all of these different presentations of the inner product can
be reduced to the standard one, as long as in the computation of
$\langle u,v \rangle$ we take the representation of vector $u$ using
the given generators $a_i$ and the representation of vector $v$
using the generators $a'_i=a_i\in \Z_2$, for $1\leq i\leq \alpha$,
and $a_i'=\bar{s_i}a_i\in \Z_4$,  for $\alpha+1\leq i\leq
\alpha+\beta$.

We can also write the standard inner product as $$\langle u,v \rangle= u {\cdot} J_n {\cdot}
v^{t},$$ where
$\dd J_n=\left (\begin{array}{c|c} 2I_{\alpha}& \zero \\
\hline \zero &I_{\beta}\end{array}\right )$ is a diagonal
matrix over $\Z_4$.
Note that when $\alpha=0$ the inner product is the usual one for
$\Z_4$-vectors (i.e. vectors over $\Z_4$) and when $\beta=0$ it is
twice the usual one for $\Z_2$-vectors.

\bigskip
Let $\codi$ be a $\add$-additive code of type
$(\alpha,\beta;\gamma,\delta;\kappa$) and let $C=\Phi(\codi)$ be the
corresponding $\add$-linear code. The {\it additive orthogonal code}
of $\codi$, denoted by ${\cal C}^\perp$, is defined in the standard
way $${\cal C}^\perp=\{v\in \Z_2^\alpha \times \Z_4^\beta \;|\;
\langle u,v \rangle =0 \mbox{ for all } u\in {\cal C}\}.$$ We will
also call ${\cal C}^\perp$ the {\it additive dual code} of ${\cal
C}$. The corresponding binary code $\Phi({\cal C}^\perp)$ is denoted
by $C_\perp$ and called {\it $\add$-dual code} of $C$. In the case
that $\alpha=0$, so when $\codi$ is a quaternary linear code, ${\cal
C}^\perp$ is also called the {\it quaternary dual code} of ${\cal
C}$ and $C_\perp$ the {\it $\Z_4$-dual code} of $C$.

The additive dual code $\mathcal{C}^\perp$ is also a
$\add$-additive code, that is a subgroup of $\Z_2^{\alpha}\times
\Z_4^{\beta}$. Its weight enumerator polynomial is related to the
weight enumerator polynomial of $\mathcal{C}$ by McWilliams
Identity (see \cite{del}).  Notice that $C$ and $C_\perp$ are not
dual in the binary linear sense but the weight enumerator
polynomial of $C_\perp$ is the McWilliams transform of the weight
enumerator polynomial of $C$ (see \cite{del}, \cite{PropIT}).

\begin{lemm}\label{mcw} Let $\codi$ be a $\add$-additive code of type
  $(\alpha,\beta;\gamma,\delta;\kappa)$ and $\codi^\perp$ its additive dual code. Then,   $|\mathcal{C}||\mathcal{C}^\perp|=2^n$, where $n=\alpha+2\beta$.
\end{lemm}

\begin{demo} ~From the McWilliams Identity, $$
    W_{\mathcal{C}^\perp}(X,Y)=\frac{1}{|\mathcal{C}|}W_{\mathcal{C}}(X+Y,X-Y).
$$ Taking $X=Y$ we obtain, $$
    |\mathcal{C}^\perp|X^n=\frac{1}{|\mathcal{C}|}(2X)^{n-wt(\zero)}
$$ and hence $|\mathcal{C}^\perp||\mathcal{C}|=2^n$. \end{demo}

Finally, notice again that one could think on $\add$-additive codes (or
$\add$-linear codes) only as quaternary linear codes (or
$\Z_4$-linear codes), changing ones by twos in the coordinates
over $\Z_2$. However, they are not equivalent to the quaternary linear code, since the inner product
defined in $\Z_2^\alpha \times \Z_4^\beta$ gives us that the dual code of a $\add$-additive code is not
equivalent to the dual code of the corresponding quaternary linear code. Take, for example,
$\alpha=\beta=1$ and the vectors $v=(1,3)$ and $w=(1,2)$. It is
easy to check that $ \langle v,w \rangle =0$, so $v$ and $w$ are
orthogonal. If we change the ones by twos in the
coordinates over $\Z_2$ of these vectors we get $v'=(2,3)$ and
$w'=(2,2)$, which are not orthogonal in the quaternary sense.

\section{Parity-check matrices of $\add$-additive codes}
\label{sec:parityCheck}

In this section, first we will prove two different methods to
construct the additive dual code of a $\add$-additive code and we
will compute the type of this additive dual code. Then, we will
apply one of these two methods to show how to construct a
parity-check matrix of a $\add$-additive code, or equivalently a
generator matrix of its additive dual code, when the $\add$-additive
code is generated by a canonical generator matrix as in
(\ref{eq:StandardForm}).

\bigskip
Let $\codi$ be a $\add$-additive code of type $(\alpha,\beta;\gamma,\delta;\kappa)$.
Since $\codi$ is a subgroup of $\Z_2^{\alpha}\times \Z_4^{\beta}$,
the code $\codi$ could be seen as the kernel of a group
homomorphism onto $\Z_2^{\bgamma}\times \Z_4^{\bdelta}$, that is,
$\codi=ker \:\vartheta$, where $$ \vartheta:\;\Z_2^{\alpha}\times
\Z_4^{\beta} \longrightarrow\;\Z_2^{\bgamma}\times \Z_4^{\bdelta}.
$$ The additive dual code $\codi^\perp$ is also the kernel of
another group homomorphism onto $\Z_2^{\gamma}\times
\Z_4^{\delta}$, that is, $\codi^\perp=ker \:{\bar \vartheta}$, where $$
{\bar \vartheta}:\;\Z_2^{\alpha}\times \Z_4^{\beta}
\longrightarrow\;\Z_2^{\gamma}\times \Z_4^{\delta}. $$

The homomorphism $\vartheta$ can be represented by a matrix $\cH$,
which can be viewed as a parity-check matrix for the
$\add$-additive code ${\cal C}$ or as a generator matrix for its
additive dual code ${\cal C}^\perp$. Vice versa, the homomorphism
$\bar \vartheta$ can be represented by a matrix $\cG$, which can be
viewed as a parity-check matrix for the additive dual code ${\cal
C}^\perp$ or as a generator matrix for the $\add$-additive code
${\cal C}$.

\begin{ex} \label{example:C1b} The code $\codi_1$ (or the corresponding
$C_1=\Phi(\codi_1)$) in Example \ref{example:C1} is a $\add$-additive code
(or a $\add$-linear code) of type $(1,3;1,2;1)$ with generator
matrix
$$\cG_1= \left (\begin{array}{c|c c c} 1 & 2 & 2 & 2\\ \hline
                                     0 & 1 & 1 & 0\\
                                     1 & 1 & 2 & 3
\end{array}\right ).$$
The generator matrix $\cG_1$ for ${\cal C}_1$ can be also viewed as a
parity-check matrix for its additive dual code ${\cal C}_1^\perp$.
Notice also that $|\codi_1|=|C_1|=2\cdot 4^2=32$, so by Lemma
\ref{mcw}, $|\codi_1^\perp|=2^7/32=4$. \end{ex}

\medskip
In order to construct the additive dual code of a $\add$-additive code,
we will need the following maps:
$\xi$ from $\Z_4$ to $\Z_2$ which is
the usual one modulo two, that is $\xi(0)=0$, $\xi(1)=1$, $\xi(2)=0$,
$\xi(3)=1$; 
and the identity map $\iota$ from $\Z_2$ to $\Z_4$, that is
$\iota(0)=0$, $\iota(1)=1$. These maps can be extended to the maps
$(\xi, Id): \Z_4^\alpha\times\Z_4^\beta \longrightarrow
\Z_2^\alpha\times\Z_4^\beta$ and $(\iota,Id):
\Z_2^\alpha\times\Z_4^\beta \longrightarrow
\Z_4^\alpha\times\Z_4^\beta$, which will also be denoted by $\xi$
and $\iota$, respectively. Recall also the map $\chi$ from $\Z_2$ to
$\Z_4$ which is the normal inclusion from the additive structure in
$\Z_2$ to $\Z_4$, that is $\chi(0)=0$, $\chi(1)=2$; and its
extension  $(\chi,Id): \Z_2^\alpha\times\Z_4^\beta \longrightarrow
\Z_4^\alpha\times\Z_4^\beta$, denoted also by $\chi$. We denote by
$\langle\cdot,\cdot\rangle_4$ the standard inner product for
quaternary vectors.

\medskip
\begin{lemm} \label{lemma:innerProduct}
If $u\in\Z_2^\alpha\times\Z_4^\beta$, $v\in\Z_4^{\alpha+\beta}$, then
$\langle \chi(u),v \rangle_4=\langle u,\xi(v)\rangle.$
\end{lemm}

\begin{demo}
$\langle \chi(u),v\rangle_4 = \sum_{i=1}^{\alpha}
(2u_i)v_i+\sum_{j=\alpha+1}^{\alpha+\beta}u_jv_j=\sum_{i=1}^{\alpha}
(2u_i)(v_i \bmod 2)+\sum_{j=\alpha+1}^{\alpha+\beta}u_jv_j= \langle
u,\xi(v) \rangle$.
\end{demo}

\begin{coro}\label{coro:chiiota}
If $u,v\in\Z_2^\alpha\times\Z_4^\beta$, then $
\langle \chi(u),\iota(v) \rangle_4=\langle u,v\rangle.$
\end{coro}

\begin{demo}
By Lemma \ref{lemma:innerProduct}, $\langle \chi(u),\iota(v)
\rangle_4=\langle u,\xi(\iota(v))\rangle=\langle u,v\rangle$.
\end{demo}

\begin{prop}\label{prop:xichiperp} Let $\cal{C}$ be a $\add$-additive
code of type $(\alpha,\beta;\gamma,\delta;\kappa)$. Then, $$
\cal{C}^\perp=\xi(\chi(\cal{C})^\perp).$$ \end{prop}

\begin{demo}
We know that if $v\in\cal{C}^\perp$, then $\langle u,v \rangle=0$, for all $u\in\cal{C}$.
By Corollary \ref{coro:chiiota}, $\langle u,v \rangle=\langle \chi(u), \iota(v) \rangle _4=0$.
Therefore, $\xi(\iota(v))=v \in \xi(\chi(\cal{C})^\perp)$ and $\cal{C}^\perp \subseteq \xi(\chi(\cal{C})^\perp)$.
On the other hand, if $v\in \chi(\codi)^\perp$, then $\langle \chi(u),v \rangle _4=0$, for all $u\in\cal{C}$.
By Lemma \ref{lemma:innerProduct}, $\langle \chi(u),v \rangle _4=\langle u,\xi(v)\rangle =0$.
Thus, $\xi(\chi(\cal{C})^\perp)\subseteq \cal{C}^\perp$ and we
obtain the equality.
\end{demo}

\begin{prop}\label{prop:xihperp} Let $\cal{C}$ be a $\add$-additive code
of type $(\alpha,\beta;\gamma,\delta;\kappa)$.
Then, $$\mathcal{C}^\perp=\chi^{-1}(\xi^{-1}(\mathcal{C})^\perp).$$
\end{prop}

\begin{demo} Let $\cG$ be a generator matrix of the $\add$-additive code
$\cal{C}$ written as in (\ref{eq:matrixG}). Then,
the quaternary linear code $\xi^{-1}(\codi)$ has a generator matrix
of the form \EQ \label{ximat} \left (\begin{array}{c c}
                2I_{\alpha} & \zero \\
                B_1 & 2B_3 \\
                B_2 & Q \\
 \end{array}\right ).\EN
 We will show that $v \in \mathcal{C}^\perp$ if and only if $ \chi(v)\in \xi^{-1}(\codi)^\perp$. In fact,
for each row vector $f$ in the matrix $(2I_{\alpha} \ \zero)$, we
have $\langle \chi(v),f \rangle _4= \sum_{i=1}^\alpha f_i 2v_i=0$
because there is only one index $i$ such that $f_i=2$. Moreover, by
Corollary \ref{coro:chiiota}, $0=\langle v,u\rangle=\langle
\chi(v),\iota(u) \rangle_4$, for all $u \in \mathcal{C}$.
\end{demo}

The following question we will settle is the computation of the type
of the additive dual code of a given $\add$-additive code
$\codi$. First, we will remember
this well-known result for quaternary linear codes, that is for $\add$-additive
codes with $\alpha=0$. Then, we will generalize it  for
$\add$-additive codes, not necessarily quaternary linear codes.

\begin{lemm}{\cite{Sole}} \label{z4}
If $\codi$ is a quaternary linear code of type $(0,\beta;\gamma,\delta;0)$,
then the quaternary dual code $\codi^\perp$ is of type
$(0,\beta;\gamma,\beta-\gamma-\delta;0)$.
\end{lemm}

\begin{theo} \label{parameters} Let
$\codi$ be a $\add$-additive code of type
$(\alpha,\beta;\gamma,\delta;\kappa)$. The additive dual code
$\codi^\perp$ is then of type $(\alpha,\beta;\bgamma,\bdelta;\bkappa)$,
where $$\begin{array}{l} \bgamma = \alpha + \gamma - 2\kappa,\\
\bdelta =\beta - \gamma - \delta + \kappa,\\
 \bkappa=\alpha-\kappa. \end{array}$$
\end{theo}

\begin{demo} Let $\cG$ be a generator matrix of the $\add$-additive code
$\cal{C}$ written as in (\ref{eq:matrixG}). Then, the matrix (\ref{ximat}) is a
generator matrix for the quaternary linear code
$\xi^{-1}(\codi)$, which is of type $(0,\alpha+\beta; \gamma',
\delta';0)$, where $\gamma'=\alpha+\gamma-2\kappa$ and
$\delta'=\delta+\kappa$. The value of $\delta'$ comes from the
fact that the $\kappa$ independent binary vectors of $(\codi_b)_X$ are in
$A$ and, so, the number of independent quaternary vectors of order
four becomes $\delta+\kappa$. The value of $\gamma'$ comes from
the fact that the cardinality of the quaternary linear code
$\xi^{-1}(\codi)$ is
$2^{\gamma'+2\delta'}=2^{\gamma+2\delta+\alpha}$.

By Lemma \ref{z4},
the quaternary dual code $\xi^{-1}(\codi)^\perp$  is of type $(0,\alpha+\beta; \bgamma, \bdelta;0)$,
where $\bgamma=\gamma'$ and $\bdelta=\alpha+\beta-\gamma'-\delta'
=\alpha+\beta-(\gamma+\alpha-2\kappa)-(\delta+\kappa)=
\beta-\gamma-\delta+\kappa$.

Note that the $\bdelta$ independent vectors in
$\xi^{-1}(\codi)^\perp$, restricted to the first $\alpha$ coordinates, are
vectors of order two, because in $\xi^{-1}(\codi)$ there are the
row vectors of the matrix $(2I_{\alpha} \ \zero)$. Finally, applying $\chi^{-1}$ we
obtain the additive dual code of $\codi$. For this additive dual code $\codi^\perp$,
the value of $\bkappa$ can be easily computed from the fact that,
again, the additive dual coincides with $\codi$.
\end{demo}

There are two different methods to obtain the additive dual code
$\codi^\perp$, one given by Proposition \ref{prop:xichiperp} and
another one by Proposition \ref{prop:xihperp}.  Using any of these
two methods, we can construct a generator matrix of $\codi^\perp$,
or equivalently a parity-check matrix of $\codi$, starting from a
generator matrix of $\codi$. In Example \ref{example:C1checkmat}, we
consider the canonical generator matrix of a $\Z_2\Z_4$-additive
code and apply these two methods to obtain a generator matrix of its
additive dual code. Note that the process to obtain this matrix is
different using both methods but, in this case, the generator
matrices obtained coincide.

Theorem \ref{prop:Dual1} shows how to construct the parity-check matrix of a $\add$-additive code
generated by a canonical generator matrix as in (\ref{eq:StandardForm}). This result is proved using the method given by Proposition \ref{prop:xichiperp}.
Notice also that we can apply any of the two methods to any generator matrix, not necessary a
canonical generator matrix, to get a parity-check matrix.

\begin{lemm} \cite{Sole} \label{lem:QaryDual1} If $\mathcal{C}$ is a quaternary linear code of type
$(0,\beta;\gamma,\delta;0)$ with canonical generator matrix
(\ref{eq:QaryStandardForm}), then the generator matrix of
$\mathcal{C}^\perp$ is  \EQ  \cH_S=\left (
\begin{array}{|ccc}
\zero                   & 2I_{\gamma}& 2R^{t}\\ \hline
I_{\beta-\gamma-\delta} & T^{t}     & -(S+RT)^{t}\\
\end{array} \right ),\EN where $R,T$ are
matrices over $\Z_2$ of size $\delta\times\gamma$ and
$\gamma\times(\beta-\gamma-\delta)$, respectively; and $S$ is a
matrix over $\Z_4$ of size $\delta\times(\beta-\gamma-\delta)$.
\end{lemm}


\begin{theo} \label{prop:Dual1} Let $\mathcal{C}$ be a $\add$-additive code of type
$(\alpha,\beta;\gamma,\delta;\kappa)$ with canonical generator matrix
(\ref{eq:StandardForm}). Then, the generator matrix of
$\mathcal{C}^\perp$ is  \EQ  \cH_S=\left (
\begin{array}{cc|ccc}
T_b^{t} & I_{\alpha-\kappa} & \zero &  \zero & 2S_b^{t}\\
\zero & \zero & \zero & 2I_{\gamma-\kappa}& 2R^{t}\\
\hline T_2^{t} &\zero & I_{\beta+\kappa-\gamma-\delta} &  T_1^{t}
& -\big( S_q+RT_1\big)^{t} \end{array} \right ),\EN
where $T_b, T_1, T_2, R, S_b$ are matrices over $\Z_2$
and $S_q$ is a matrix over $\Z_4$.
\end{theo}

\begin{demo} By Lemma \ref{lem:QaryDual1}, if $\bar{\codi}$ is a quaternary linear code
with generator matrix $$\bar{\cG}= \left ( \begin{array}{|ccccc}
2T_b & 2T_2 & 2I_{\kappa} & \zero & \zero\\
\zero & 2T_1 & \zero & 2I_{\gamma-\kappa} & \zero\\ \hline
2S_b & S_q & \zero & R & I_{\delta} \end{array} \right ),$$ then
the quaternary dual code $\bar{\codi}^\perp$ has generator matrix $\bar{\cH}=$
$$ \left ( \begin{array}{|ccccc}
\zero & \zero & 2I_{\kappa} & \zero & \zero \\
\zero & \zero & \zero & 2I_{\gamma-\kappa}& 2R^{t}\\ \hline
I_{\alpha-\kappa} & \zero & T_b^{t} & \zero & 2S_b^{t}\\
\zero & I_{\beta-\gamma-\delta+\kappa} & T_2^{t} & T_1^{t} &
-\big( S_q+RT_1\big)^{t} \end{array} \right ). $$
Hence, if $\codi$ is a $\add$-additive code with generator
matrix (\ref{eq:StandardForm}), then the generator matrix of
$\chi(\codi)^\perp$ is $\cH_\xi=$  $$\left (
\begin{array}{|ccccc}
2I_{\kappa} &\zero & \zero &  \zero & \zero \\
\zero & \zero & \zero & 2I_{\gamma-\kappa}& 2R^{t}\\ \hline
T_b^{t} & I_{\alpha-\kappa} & \zero &  \zero & 2S_b^{t}\\
T_2^{t} &\zero & I_{\beta-\gamma-\delta+\kappa} &  T_1^{t} &
-\big( S_q+RT_1\big)^{t} \end{array} \right ). $$
Finally, by Proposition \ref{prop:xichiperp}, $\cH_S=\xi(\cH_\xi)$ is
the generator matrix of $\codi^\perp$. \end{demo}

Note that by Theorem \ref{parameters} and Theorem \ref{prop:Dual1},
if $\mathcal{C}$ is a $\add$-additive code of type
$(\alpha,\beta;\gamma,\delta;\kappa)$ with canonical generator matrix
(\ref{eq:StandardForm}), then $\codi^\perp$ is permutation equivalent
to a $\add$-additive code with canonical generator matrix
\EQ  \left ( \begin{array}{cc|ccc}
I_{\bkappa} & T_b^{t} & 2S_b^{t} & \zero &  \zero \\
\zero & \zero & 2R^{t} & 2I_{\bgamma-\bkappa}& \zero \\ \hline
 \zero & T_2^{t} & -\big( S_q+RT_1\big)^{t} &  T_1^{t} &
I_{\bdelta} \end{array} \right ), \EN where $T_b, T_1, T_2, R, S_b$ are matrices over $\Z_2$;
$S_q$ is a matrix over $\Z_4$, $\bgamma=\alpha+\gamma-2\kappa$,
$\bdelta=\beta-\gamma-\delta+\kappa$ and $\bkappa=\alpha-\kappa$.

\medskip
\begin{ex} \label{example:C1checkmat}
Let ${\codi_{S1}}$ denote the $\add$-additive code of type $(1,3;1,2;1)$ with canonical
generator matrix
$$\cG_S=\left (\begin{array}{c|ccc} 1 & 2 & 0 & 0\\ \hline
                                     0 & 1 & 1 & 0\\
                                     0 & 3 & 0 & 1
\end{array}\right ).$$
By Theorem \ref{parameters}, the additive dual code $\codi_{S1}^\perp$ is of type $(1,3;0,1;0)$.
There are two methods to obtain a parity-check matrix of ${\codi_{S1}}$ from the matrix
$\cG_S$.

The first one uses Proposition \ref{prop:xichiperp}.
We know that if $\bar\codi$ is a quaternary linear code with generator matrix
$\bar{\cG} =\left (\begin{array}{|cccc} 2 & 2 & 0 & 0\\ \hline
                            1 & 0 & 1 & 0\\
                            3 & 0 & 0 & 1
\end{array}\right ),$ the quaternary dual code $\bar\codi^\perp$ has generator matrix
$\bar\cH =\left (\begin{array}{|c ccc} 0 & 2 & 0 & 0\\  \hline
                            1 & 1 & 3 & 1 \\
\end{array}\right ).$ So, the generator matrix of $\chi(\codi_{S1})^\perp$ is
$\left (\begin{array}{|c ccc} 2 & 0 & 0 & 0\\ \hline
                            1 & 1 & 3 & 1 \\
\end{array}\right )$
and finally, applying $\xi$, the generator matrix of
$\codi_{S1}^\perp=\xi(\chi(\codi_{S1})^\perp)$
is $${\cH_S} =\left (\begin{array}{c|ccc}  1 & 1 & 3 & 1\\
\end{array}\right ).$$

The second method uses Proposition \ref{prop:xihperp}. We know that the quaternary
linear code $\xi^{-1}(\codi_{S1})$ with generator matrix
$$\left (\begin{array}{|c c c c} 2 & 0 & 0 & 0 \\ \hline
                                  1 & 2 & 0 & 0 \\
                                  0 & 1 & 1 & 0 \\
                                  0 & 3 & 0 & 1 \\
                                   \end{array}\right ),$$
or equivalently $\left (\begin{array}{|c c c c} \hline
                                  1 & 2 & 0 & 0 \\
                                  0 & 1 & 1 & 0 \\
                                  0 & 3 & 0 & 1 \\
                                   \end{array}\right )$,
has parity-check matrix $\left (\begin{array}{|cccc} \hline  2 & 1 & 3 & 1\\
\end{array}\right ).$ So, applying $\chi^{-1}$, the generator matrix of
$\codi_{S1}^\perp=\chi^{-1}(\xi^{-1}(\codi_{S1})^\perp)$
is $${\cH_S} =\left (\begin{array}{c|ccc}  1 & 1 & 3 & 1\\
\end{array}\right ).$$
\end{ex}

\begin{ex} Let ${\codi_{S2}}$ be a $\add$-additive code of type $(3,4;3,1;1)$ with canonical generator
matrix
$$\left (\begin{array}{ccc|cccc}1 & 0 & 0  &  2 & 2 & 0 & 0 \\
                                0 & 1 & 0  &  0 & 0 & 0 & 0 \\
                                0 & 0 & 1  &  2 & 2 & 0 & 0 \\\hline
                                0 & 0 & 0  &  1 & 1 & 1 & 1 \\
\end{array}\right ).$$
By Theorem \ref{parameters} and Theorem \ref{prop:Dual1}, the additive dual
code $\codi_{S2}^\perp$ is of type $(3,4;0,3;0)$ and has generator matrix
$$\left (\begin{array}{ccc|cccc}\hline 1 & 0 & 1  &  1 & 0 & 0 & 3 \\
                                       1 & 0 & 1  &  0 & 1 & 0 & 3 \\
                                       0 & 0 & 0  &  0 & 0 & 1 & 3 \\
\end{array}\right ).$$
\end{ex}

\section{Additive self-dual codes}
\label{sec:selfdual}

Let $\codi$ be a $\add$-additive code. We say that $\codi$ is an {\it additive self-orthogonal code} if $\codi\subseteq \codi^\perp$
and $\codi$ is an {\it additive self-dual code} if $\codi= \codi^\perp$.
Let $C=\Phi(\codi)$ be the corresponding
$\add$-linear code. We say that $C$ is a {\em self\/}
$\add$-{\em orthogonal code} if $C\subseteq C_\perp$ and $C$ is a {\em
self\/} $\add$-{\em dual code} if $C=C_\perp$, where $C_\perp=\Phi(\codi^\perp)$.
In this section, we will study the additive self-dual codes.

\bigskip
Note that in the case that $\beta=0$, that is when $\codi=C$ is a binary
linear code, we will also say that $\codi$ is {\em binary
self-orthogonal} (or {\em binary self-dual}) if $\codi \subseteq
\codi^\perp$ (or $\codi= \codi^\perp$). And in the case that
$\alpha=0$, that is when $\codi$ is a quaternary linear code, we will
also say that $\codi$ is {\em quaternary self-orthogonal} (or {\em
quaternary self-dual}) if $\codi \subseteq \codi^\perp$ (or $\codi=
\codi^\perp$).

Recall that $\codi_X$ is the punctured code of $\codi$ by deleting
the coordinates outside $X$, $\codi_Y$ is the punctured code of
$\codi$ by deleting the coordinates outside $Y$ and $\codi_b$ is the
subcode of $\codi$ which contains all codewords of order two.
Denote by $w(u)$ the Hamming weight of any vector $u\in \Z_2^\alpha$.

\bigskip
\begin{lemm}\label{sd:parameters}
If $\codi$ is an additive self-dual code, then $\codi$ is of type
$(2\kappa,\beta;\beta+\kappa-2\delta,\delta;\kappa)$,
$|\codi|=2^{\kappa+\beta}$ and $|\codi_b|=2^{\kappa+\beta-\delta}$.
\end{lemm}

\begin{demo}
By Theorem \ref{parameters}, we have that $\alpha=2\kappa$ and
$\gamma=\beta+\kappa-2\delta$. Since $|\codi|=2^{\gamma+2\delta}$ and
$|\codi_b|=2^{\gamma+\delta}$, the result holds.
\end{demo}

\begin{lemm}\label{sd:evenodd}
Let $\codi$ be an additive self-dual code and let $z=(x\mid y)\in
\codi$. Denote by $p(u)$ the number of odd (order four) coordinates
of any vector $u\in \Z_4^\beta$. Then,
\begin{itemize}
\item[(i)] if $w(x)$ is even, then $p(y)\equiv 0 \pmod 4$.
\item[(ii)] if $w(x)$ is odd, then $p(y)\equiv 2 \pmod 4$.
\item[(iii)] $(\zero\mid \mathbf{2})$ is a codeword in $\codi$.
\end{itemize}
\end{lemm}

\begin{demo}
$(i)$ and $(ii)$ follows easily since $z$ must be orthogonal to
itself and we have $\langle z,z \rangle=2w(x)+p(y)=0 \in \Z_4$. Now,
$(iii)$ is obvious because $p(y)$ is always even.
\end{demo}

\begin{lemm}\label{sd:binariautodual}
If $\codi$ is an additive self-dual code,
then the subcode $(\codi_b)_X$ is a binary self-dual code.
\end{lemm}

\begin{demo}
By Lemma \ref{sd:parameters}, the code $\codi$ is of type $(2\kappa,\beta;\beta+\kappa-2\delta,\delta;\kappa)$.
Since for any pair of codewords $(x\mid y), (x'\mid
y')\in \codi_b$ we have $\langle y,y' \rangle_4=0$, $(\codi_b)_X\subseteq (\codi_b)^\perp_X$.
Moreover, since $(\codi_b)_X$ has dimension $\kappa$
(by definition) and is of length $2\kappa$, we have that $(\codi_b)_X$ is
binary self-dual.
\end{demo}

\begin{lemm}\label{sd:cardinalquaternari}
Let $\codi$ be an additive self-dual code of type $(2\kappa,\beta; \beta+\kappa-2\delta,\delta;\kappa)$.
There is an integer number $r$, $0\leq r\leq \kappa$, such that each
codeword in $\codi_Y$ appears $2^r$ times in $\codi$ and
$|\codi_Y|\geq 2^\beta$.
\end{lemm}

\begin{demo}
Consider the subcode $\codi_0=\{(x\mid \zero)\in \codi\}$. Clearly,
$(\codi_0)_X$ is a binary linear code. Let $r=dim(\codi_0)_X$. Thus,
any vector in $\codi_Y$ appears $2^r$ times in $\codi$. Note that
$(\codi_0)_X$ is also a subcode of $(\codi_b)_X$, hence $r\leq \kappa$.
Also, we have that $|\codi|=2^{\beta +\kappa}=|\codi_Y|\cdot 2^r$,
therefore $|\codi_Y|\geq 2^\beta$.
\end{demo}

We say that a binary code $C$ is {\it antipodal} if for any codeword
$z\in C$, $z+\u \in C$. The following two examples show us two
different cases of additive self-dual codes. In Example
\ref{sd:antipodal}, the corresponding $\add$-linear code
$C_1=\Phi(\codi_1)$ is antipodal, or equivalently $\codi_1$ contains
the codeword $(\u\mid \mathbf{2})$. On the other hand, in Example
\ref{sd:noantipodal}, the corresponding $\add$-linear code
$C_2=\Phi(\codi_2)$ is not antipodal. We will study these two cases
separately.

\begin{ex}\label{sd:antipodal}
An additive self-dual code with $\alpha,\beta \geq  1$ should have
$\alpha\geq 2$, since $\alpha$ must be even. An additive self-dual
code with minimum number of coordinates has $\alpha=2$, $\beta=1$
and $2^{\kappa+\beta}=2^{1+1}=4$ codewords. For example, the code
$\codi_1=\{(00\mid 0), (00\mid 2), (11\mid 0), (11\mid 2)\}$ is an
additive self-dual code of type $(2,1;2,0;1)$ and has generator
matrix
$$\cG_1=\left (\begin{array}{cc|c}    1  &  1 & 0 \\
                                0  &  0 & 2 \\\hline
\end{array}\right ).$$
\end{ex}

\begin{ex}\label{sd:noantipodal}
Let $\codi_b=\{(00\mid 00), (00\mid 22), (11\mid 02), (11\mid 20)\}$.
Then, the code $\codi_2=\codi_b \cup (\codi_b+(01\mid 11))$ is an additive
self-dual code of type $(2,2;1,1;1)$ and has generator matrix
$$\cG_2=\left (\begin{array}{cc|cc}  1 & 1  &  2 & 0 \\ \hline
                                0  &1 &  1 & 1 \\
\end{array}\right ).$$
\end{ex}

The following Lemmas \ref{lemEx1} and \ref{lemEx2} give us two generalizations of Example \ref{sd:noantipodal}. Note that any of the corresponding $\add$-linear codes are not antipodal.

\begin{lemm} \label{lemEx1} If $\delta \leq \kappa$, the $\add$-additive code $\codi$ of type $(2\kappa,\beta;\beta+\kappa-2\delta,\delta;\kappa)$
with canonical generator matrix $$\cG=\left (\begin{array}{cccc|ccc}
    I_\delta  & \zero & I_\delta & \zero & 2I_\delta & \zero & \zero \\
       \zero  & I_{\kappa-\delta}& \zero & I_{\kappa-\delta} & \zero & \zero & \zero \\
       \zero & \zero &  \zero  &  \zero & \zero & 2I_{\beta-2\delta} & \zero \\ \hline
       \zero  & \zero & I_\delta & \zero & I_\delta & \zero & I_\delta \\
\end{array}\right )$$ is an additive self-dual code.
\end{lemm}

\begin{demo} Straightforward using Theorem \ref{prop:Dual1}. \end{demo}

\begin{lemm} \label{lemEx2} If $\delta \leq \kappa$,
the $\add$-additive code $\codi$ of type $(2\kappa,\beta;\beta+\kappa-2\delta,\delta;\kappa)$
with canonical generator matrix $$\cG=\left (\begin{array}{cccc|ccc}
    I_\delta  & \zero & I_\delta & \zero & 2I_\delta & \zero & \zero \\
       \zero  & I_{\kappa-\delta}& \zero & I_{\kappa-\delta} & \zero & \zero & \zero \\
       \zero & \zero &  \zero  &  \zero & \dos & 2I_{\beta-2\delta} & \zero \\ \hline
       \zero  & \zero & I_\delta & \zero & I_\delta & \u & I_\delta \\
\end{array}\right )$$ is an additive self-dual code if and only if $\beta -2\delta \equiv 0 \pmod 4$.
\end{lemm}

\begin{demo} Let $e_i$ denote the vector  with all components equal to zero, except the
$i$th component, which contains a one. It is easy to  that any two
rows of the generator matrix $\cG$ are orthogonal. Notice that the
rows of order four $u=(\zero \ \zero \ e_i \ \zero \mid e_i \ \u \
e_i)$  are orthogonal if and only if $\beta -2\kappa \equiv 0\pmod
4$. By Theorem \ref{prop:Dual1}, we have $|\codi|=|\codi^\perp|$, so
$\codi$ is additive self-dual.
\end{demo}

\begin{prop}\label{sd:propantipodal}
Let $C$ be a self $\;\add$-dual code of length $n$ and let $A_i$
denote the number of codewords of weight $i$ ($0\leq i\leq n$). The
following statements are equivalent:
\begin{itemize}
\item[(i)] $C$ is antipodal.
\item[(ii)] $C_X$ has only even weights (and also $C$).
\item[(iii)] $\sum_{i=0}^n (-1)^i A_i=|C|$.
\end{itemize}
\end{prop}

\begin{demo}
Let $\codi$ be the corresponding $\add$-additive code $\codi=\Phi^{-1}(C)$.
Note that given a codeword $z=(x\mid y)\in \codi$, where
$y=(y_1,\ldots,y_\beta)$, the weight of $(\phi(y_1),\ldots,\phi(y_\beta))$ is always even.
Thus, the parity of the weight of $\Phi(z)$
is the same as the parity of the weight of $x$.

$(i)\;\Leftrightarrow\;(ii)$: If $C$ is antipodal, then $(\u \mid
\mathbf{2})\in \codi$. Therefore, for any codeword $(x\mid y)\in
\codi$, $w(x)$ must be even. Reciprocally, if $w(x)$ is even for any
codeword $(x\mid y)\in \codi$, then the vector $(\u \mid
\mathbf{2})$ is orthogonal to any codeword and hence $(\u \mid
\mathbf{2})\in \codi$.

$(ii)\;\Leftrightarrow\;(iii)$: Straightforward because
$\sum_{i=0}^n A_i=|C|$.
\end{demo}

\begin{prop}
Let $C$ be a self $\;\add$-dual code of length $n$ and let $A_i$ denote
the number of codewords of weight $i$ ($0\leq i\leq n$).
The following statements are equivalent:
\begin{itemize}
\item[(i)] $C$ is not antipodal.
\item[(ii)] $C_X$ has even and odd weights (and also $C$).
\item[(iii)] $\sum_{i=0}^n (-1)^i A_i=0$.
\end{itemize}
\end{prop}

\begin{demo}
Statements $(i)$ and $(ii)$ are equivalent by Proposition
\ref{sd:propantipodal}.

$(i)\;\Leftrightarrow\;(iii)$: Consider the MacWilliams Identity:
$$
W_C(X,Y)=\frac{1}{|C|} W_{C_\perp}(X+Y,X-Y),
$$
where $W_C(X,Y)$ is the weight enumerator polynomial of $C$:
$$
W_C(X,Y)=\sum_{i=0}^n A_i X^{n-i} Y^i.
$$
Since $C=C_\perp$ and taking $X=0$, we obtain:
$$
A_n Y^n = \frac{1}{|C|}\sum_{i=0}^n (-1)^i A_i
Y^n\;\;\Longrightarrow\;\;|C|A_n=\sum_{i=0}^n (-1)^i A_i.
$$
Finally, since $A_n=1$ when $C$ is antipodal and $A_n=0$ when $C$ is not
antipodal, we have that $\sum_{i=0}^n (-1)^i A_i=0$ if and only if
$C$ is not antipodal.
\end{demo}

\begin{prop} \label{prop:antipodal}
Let $\codi$ be an additive self-dual code of
type $(2\kappa,\beta; \beta+\kappa-2\delta,\delta;\kappa)$. The following statements
are equivalent:
\begin{itemize}
\item[(i)] $\codi_X$ is binary self-orthogonal.
\item[(ii)] $\codi_X$ is binary self-dual.
\item[(iii)] $|\codi_X|=2^\kappa$.
\item[(iv)] $\codi_Y$ is a quaternary self-orthogonal code.
\item[(v)] $\codi_Y$ is a quaternary self-dual code.
\item[(vi)] $|\codi_Y|=2^\beta$.
\item[(vii)] $\codi = \codi_X\oplus\codi_Y$.
\end{itemize}
\end{prop}

\begin{demo}
$(i)\;\Leftrightarrow\;(ii)$: By Lemma \ref{sd:binariautodual},
$|\codi_X|\geq 2^\kappa$, thus $(i)$ and $(ii)$ are equivalent
statements.

$(ii)\;\Leftrightarrow\;(iii)$: Clearly, $(ii)$ implies $(iii)$ and
$(iii)$ implies $\codi_X=(\codi_b)_X$ and  $\codi_X$
is binary self-dual, by Lemma \ref{sd:binariautodual}.

$(ii)\;\Leftrightarrow\;(v)$: Straightforward.

$(iv)\;\Leftrightarrow\;(v)$: By Lemma \ref{sd:cardinalquaternari},
$|\codi_Y|\geq 2^\beta$, thus $(iv)$ and $(v)$ are equivalent
statements.

$(ii)\;\Leftrightarrow\;(vii)$: If $\codi_X$ is binary self-dual, then
$\codi_Y$ is quaternary self-dual, $|\codi_X|=2^\kappa$ and
$|\codi_Y|=2^\beta$. Since $\codi=2^{\kappa+\beta}$ we have that the
set of codewords in $\codi$ is $\codi_X \times \codi_Y$.
Reciprocally, if $\codi = \codi_X\oplus\codi_Y$, then $(x\mid
\zero)\in \codi$ for any $x\in\codi_X$ and $\codi_X$ must be a
binary self-dual code. Also, $(\zero\mid y)\in\codi$ for any $y\in\codi_Y$ and
$\codi_Y$ must be a quaternary self-dual code.

$(v)\;\Rightarrow\;(vi)$: Trivial.

$(vi)\;\Rightarrow\;(iii)$: By Lemma \ref{sd:cardinalquaternari},
each vector in $\codi_Y$ appears $2^\kappa$ times in $\codi$. Thus,
for any vector $x_b\in (\codi_b)_X$, the vector $(x_b\mid\zero)$ is a codeword
in $\codi_b$. This means that given any codeword $(x\mid y)\in\codi$, we
have that $\langle x,x_b \rangle=0$, for all $x_b\in(\codi_b)_X$, since $ \langle (x\mid
y),(x_b,\zero)\rangle = 0$. Therefore, for all $x\in\codi_X$, $x\in (\codi_b)_X^\perp$
and $\codi_X \subseteq (\codi_b)_X^\perp$. By Lemma \ref{sd:binariautodual},
$(\codi_b)_X^\perp=(\codi_b)_X$, which implies $\codi_X=(\codi_b)_X$,
hence $|\codi_X|=2^\kappa$.
\end{demo}

It is easy to check that if $\codi$ is an additive self-dual code, then the
codewords in $\codi_X^\perp\oplus\codi_Y^\perp$ are orthogonal to
$\codi$ and, hence, $\codi_X^\perp\oplus\codi_Y^\perp\subseteq
\codi^\perp$.

\begin{prop}
  If $\codi_X$ is a binary self-dual code of length $\alpha=2\kappa$
  and $\codi_Y$ is a quaternary self-dual code of type
  $(0,\beta;\gamma,\delta;0)$, then
  $\codi=\codi_X\oplus\codi_Y$ is an additive self-dual code of type
  $(2\kappa,\beta;\beta+\kappa-2\delta,\delta;\kappa)$.
\end{prop}

\begin{demo} Since $\codi_X$ is binary self-dual, $|\codi_X|=2^\kappa$, where $\alpha=2\kappa$.
Since $\codi_Y$ is quaternary self-dual, $|\codi_Y|=2^\beta=2^{\gamma+2\delta}$,
  so $\gamma=\beta-2\delta$.
  Let $\cG_X$ be a generator matrix of $\codi_X$ of size $\kappa\times 2\kappa$
  and let $\cG_Y$ be a  generator
  matrix of $\codi_Y$ of size $(\beta-\delta) \times \beta$.  Then, the matrix $\cG$ defined as
$$\cG=\left (\begin{array}{c|c} \cG_X & \zero \\
                            \zero & \cG_Y
\end{array}\right )$$
\noindent is a generator matrix of $\codi=\codi_X\oplus\codi_Y$. It
is easy to check that $\codi$ is an additive self-dual code of type
$(2\kappa,\beta;\beta+\kappa-2\delta,\delta;\kappa)$ and $|\codi|=2^{\kappa+\beta}$.
\end{demo}

Clearly, any of the statements $(i)-(vii)$ of Proposition
\ref{prop:antipodal} implies that $C=\Phi(\codi)$ is antipodal. We
are going to see that the converse is not true.

\begin{lemm}\label{sd:minim4}
Let $\codi \subset \ad$ ($\alpha,\beta \geq 1$) be an additive self-dual
code such that $C=\Phi(\codi)$ is antipodal (i.e.
$(\u\mid\dos)\in\codi$) and $\codi_X$ is not self-dual, then
$\alpha\geq 4$ and $\beta \geq 4$.
\end{lemm}

\begin{demo}
Recall that $\alpha$ must be even for an additive self-dual code. We are
assuming that $(\u\mid\dos)\in\codi$ and by Lemma \ref{sd:evenodd}
$(\zero\mid\dos)\in\codi$, thus $(\u\mid\zero)\in\codi$ implying
that any codeword has even weight in its binary coordinates.

If $\alpha=2$, then $\codi_X=\{(0,0),(1,1)\}$, which is self-dual.
Therefore $\alpha\geq 4$.

Let $x=(x_b\mid x_q)$ and $y=(y_b \mid y_q)$ be two codewords such
that $x_b$ and $y_b$ are not orthogonal. Then $x_q$ and/or $y_q$
must have order 4; otherwise $x$ and $y$ would not be orthogonal.
Assume that $x_q$ has order 4. Since $w(x_b)$ is even, then $x_q$
has at least 4 coordinates of order 4, by Lemma \ref{sd:evenodd}.
Hence $\beta\geq 4$.
\end{demo}

If we assume that such an additive self-dual code $\codi$ of type
$(4,4;\gamma,\delta;2)$ exists, the
sum of two codewords of order 4 always gives an order 2 codeword.
Hence it will have the same number of order 2 and order 4 codewords.
Thus
$$
2^{\gamma+\delta}=\frac{1}{2}\cdot 2^{\gamma+2\delta}
$$
and we obtain
that $\delta=1$. By Lemma \ref{sd:parameters}, we have
$\gamma + 2\delta = \beta + \kappa = 4+2=6,$ which
implies that $\gamma=4$.
Effectively such a code $\codi$ exists. A generator matrix for
$\codi$ is
$$
\left (\begin{array}{cccc|cccc}
1 & 0 & 1 & 0 & 2 & 0 & 0 & 0 \\
0 & 1 & 0 & 1 & 2 & 0 & 0 & 0 \\
0 & 0 & 0 & 0 & 2 & 2 & 0 & 0 \\
0 & 0 & 0 & 0 & 2 & 0 & 2 & 0 \\
\hline
0 & 0 & 1 & 1 & 1 & 1 & 1 & 1 \\
\end{array}\right ).
$$
Therefore, we have proven the following result:

\begin{prop}
An additive self-dual code $\codi$ with minimum cardinality and
number of coordinates such that $C=\Phi(\codi)$ is antipodal and
$\codi_X$ is not self-dual is of type $(4,4;4,1;2)$.
\end{prop}

\bigskip
The following lemma give us a family of additive self-dual codes $\codi$
such that $C=\Phi(\codi)$ is antipodal and $\codi_X$ is not
self-dual.

\begin{lemm} If $\delta < \kappa$, the $\add$-additive code $\codi$ of type $(2\kappa,\beta;\beta+\kappa-2\delta,\delta;\kappa)$
with canonical generator matrix $${\footnotesize \cG=\left (\begin{array}{cccccc|ccc}
    I_\delta  & \zero & \zero & I_\delta & \zero & \zero & 2I_\delta & \zero & \zero \\
        \zero & 1     & \zero & \zero    & 1     & \zero & {\bf 2} & \zero & \zero \\
       \zero  & \zero & I_{\kappa-\delta-1}& \zero & \zero & I_{\kappa-\delta-1} & \zero & \zero & \zero \\
       \zero & \zero & \zero & \zero &  \zero  &  \zero & {\bf 2} & 2I_{\beta-2\delta} & \zero \\ \hline
       \zero  & \zero & \zero & I_\delta & {\bf 1} & \zero & I_\delta & \u & I_\delta \\
\end{array}\right )}$$ is an additive self-dual code if and only if $\beta -2\delta \equiv 2 \pmod 4$.
Moreover, $C=\Phi(\codi)$ is antipodal and $\codi_X$ is not self-dual.
\end{lemm}

\begin{demo}
Using the same arguments as in the proof of Lemma \ref{lemEx2}, we
have that $\codi$ is an additive self-dual code if and only if
$\beta -2\delta \equiv 2 \pmod 4$. By Proposition
\ref{sd:propantipodal}, $C=\Phi(\codi)$ is antipodal. And, it is
clear that $\codi_X$ is not self-dual.
\end{demo}

\end{document}